\documentclass[a4paper,fleqn,usenatbib]{mnras}
\usepackage{newtxtext,newtxmath}

\usepackage[T1]{fontenc}
\usepackage{ae,aecompl}

\usepackage{graphicx}	
\usepackage{amsmath}	
\usepackage{threeparttable}
\usepackage{hyperref}
\usepackage{rotating}
\usepackage{subfigure}

\title[Major mechanism to drive turbulence]{The major mechanism to drive turbulence in star-forming galaxies}

\author[X. L. Yu et al.]{Xiaoling Yu,$^{1,2,3}$\thanks{E-mail: xiaoling@smail.nju.edu.cn}
Fuyan Bian,$^{3}$\thanks{E-mail: fbian@eso.org}
Mark R. Krumholz,$^{4,5}$
Yong Shi,$^{1,2}$
Songlin Li,$^{1,2}$
\newauthor
Jianhang Chen$^{1,6}$
\\
$^{1}$School of Astronomy and Space Science, Nanjing University, Nanjing 210093, China\\
$^{2}$Key Laboratory of Modern Astronomy and Astrophysics (Nanjing University), Ministry of Education, Nanjing 210093, China\\
$^{3}$European Southern Observatory, Alonso de Córdova 3107, Casilla 19001, Vitacura, Santiago 19, Chile\\
$^{4}$Research School of Astronomy \& Astrophysics, Australian National University, Canberra, ACT 2611, Australia\\
$^{5}$Centre of Excellence for All Sky Astrophysics in Three Dimensions (ASTRO-3D), Australia\\
$^{6}$European Southern Observatory, Karl-Schwarzschild-Strasse 2, D-85748 Garching bei Muenchen, Germany\\
}

\date{Accepted XXX. Received YYY; in original form ZZZ}

\pubyear{2021}

\begin{document}
\label{firstpage}
\pagerange{\pageref{firstpage}--\pageref{lastpage}}
\maketitle

\begin{abstract}
Two competing models, gravitational instability-driven transport and stellar feedback, have been proposed to interpret the high velocity dispersions observed in high-redshift galaxies. We study the major mechanisms to drive the turbulence in star-forming galaxies using a sample of galaxies from the xCOLD GASS survey, selected based on their star-formation rate (SFR) and gas fraction to be in the regime that can best distinguish between the proposed models. We perform Wide Field Spectrograph (WiFeS) integral field spectroscopic (IFS) observations to measure the intrinsic gas velocity dispersions, circular velocities and orbital periods in these galaxies. Comparing the relation between the SFR, velocity dispersion, and gas fraction with predictions of these two theoretical models, we find that our results are most consistent with a model that includes both transport and feedback as drivers of turbulence in the interstellar medium. By contrast, a model where stellar feedback alone drives turbulence under-predicts the observed velocity dispersion in our galaxies, and does not reproduce the observed trend with gas fraction. These observations therefore support the idea that gravitational instability makes a substantial contribution to turbulence in high redshift and high SFR galaxies.
\end{abstract}




\section{Introduction}
High redshift star-forming galaxies have higher star formation rates (SFRs) than low redshift star-forming galaxies. The SFR density peaked around the redshift range of $z=1-3$ \citep[e.g.,][]{Madau96,Lilly96,Madau14}. High redshift star-forming galaxies also differ from local counterparts in their small size, high gas fraction and clumpy, thick star-forming discs \citep[e.g.][]{Genzel11}. One of the most intriguing properties of high-redshift disc galaxies is that their gas velocity dispersions are higher by factors of 2-5 compared to local star-forming galaxies \citep[e.g.][]{Forster06,Forster09,Cresci09,Wisnioski11,Wisnioski15,Ubler19}. These high velocity dispersions likely indicate a highly turbulent ionised interstellar medium (ISM) \citep[e.g.][]{Law07,Forster09,Green14,Krumholz16,Simons17}. However, high velocity dispersions do not correlate solely with redshift. Instead, samples of large numbers of galaxies show that the gas velocity dispersion is well correlated with galactic SFR \citep[e.g.][]{Lehnert09,Lehnert13,Green10,Green14,Moiseev15,Johnson18,Yu19}.

There are two types of theoretical models to explain the high velocity dispersions in high-redshift galaxies. One type suggests that the high velocity dispersion is caused by stellar feedback, which injects more energy to the ISM per unit mass in high-redshift galaxies due to their higher SFRs \citep[e.g.][]{Faucher-Giguere13,Hayward17,Orr20}. The other type of model is to attribute the high velocity dispersion to gravitational instability \citep[e.g.][]{Bournaud07,Bournaud09,Bournaud10,Ceverino10,Goldbaum15,Goldbaum16,Krumholz16,Krumholz18}. These models suggest that gravitational instability produces non-axisymmetric torques that move mass inward, driving turbulence in the process. The process both provides fuel that prevents star formation in galactic centres from exhausting the gas supply in much less than a Hubble time, and regulates the Toomre $Q$ parameter to $Q \sim 1$ \citep{Forbes12, Forbes14, Goldbaum16, Krumholz18}. Both types of models predict a positive correlation between SFR and gas velocity dispersion. However, they are different in the details of this correlation, in particular, how the velocity dispersion increases depending on the SFR, as well as other secondary parameters, including gas fraction and circular velocity \citep{Krumholz16,Krumholz18}. Combining the observations with theoretical models, \citet{Krumholz16} argue that stellar feedback alone cannot explain the high velocity dispersions. They propose that for galaxies with high SFR and high velocity dispersion, the gravity-driven model agrees better with observations. \citet{Krumholz18} extend this work by proposing a unified model that can link mass transport, star formation fuelling and gravitational instability with the energy and momentum balance of star formation feedback. By comparing their models with collected observation data from the literature, they suggest that transport+feedback can explain the observed velocity dispersions for both local and high $z$ galaxies. The model predicts that transport dominates the turbulence of high $z$ star forming galaxies and feedback dominates the turbulence of low SFR star forming galaxies.

While these results are suggestive, the data available in the \citet{Krumholz16} and \citet{Krumholz18} studies were limited and heterogeneous. It is therefore of interest to carry out higher precision tests using more targeted and homogeneous data sets. There are two main ways to approach this task. One is using only $\rm{H\alpha}$ data, the approach followed by most authors to date \citep[e.g.,][]{Johnson18,Ubler19,Varidel20}. The advantage of this approach is the large sample size one can obtain if only H$\alpha$ data are required. The disadvantage is there is no information about molecular and atomic gas, and thus one cannot test any model predictions that depend on gas fraction or other gas properties. The alternative approach that we pursue here is to accept smaller sample sizes, but make use of data on the neutral interstellar gas. The GASS and xCOLD GASS survey provide well-measured molecular gas and atomic gas masses based on CO and H~\textsc{i} 21 cm observations \citep{Saintonge11,Saintonge12,Saintonge17,Catinella12}. In this work, we select a representative sample of star-forming galaxies based on their SFRs and molecular and atomic gas masses from the xCOLD GASS survey (Section \ref{sec:2-1}).
We carry out IFS observations of these star-forming galaxies. Such observations provide robust measurements of velocity dispersion, circular velocity, and orbital period in these galaxies, which can be used to better test these two theoretical models. In Section \ref{sec:2}, we describe the details of sample selection, observations, data reduction and data analysis. Section \ref{sec:5} shows discusses the results and their implications. We summarise our main conclusions in Section \ref{sec:6}. We use the cosmological parameters $H_0$ = 70 $\rm{km\ s^{-1}\ Mpc^{-1}}$, $\Omega_m$ = 0.3, $\Omega_{\Lambda}$ = 0.7 throughout this paper.


\section{Observations, Data Reduction, data analysis and theoretical models}
\label{sec:2}
\subsection{Sample Selection}
\label{sec:2-1}
We draw our sample of galaxies from the xCOLD GASS survey, selecting targets with SFRs and gas fractions in the regime that provides the greatest sensitivity to whether turbulence is driven primarily by gravity or by feedback. These galaxies have typical gas fraction of $f_{\rm g} \simeq$ 0.3 (where $f_{\rm g} \equiv M_{\rm g} /(M_{\rm g} +  M_{*})$, for gas mass $M_{\rm g}$ and stellar mass $M_*$) and SFR in the range $4\ \rm M_\odot \ yr ^{-1}$ to $13\ \rm M_\odot \ yr ^{-1}$. We focus on this range because \citet{Krumholz18} show that for this range of gas fraction and SFR, the feedback-only and feedback plus gravity models make very different predictions: the feedback plus gravity model predicts both higher velocity dispersion and a strong relationship between gas fraction and velocity dispersion that is absent in the feedback-only model. Thus both the absolute value of the velocity dispersion and the scaling between velocity dispersion and gas fraction provide means of distinguishing between the models in this regime. We exclude the galaxies with close companions to minimize the impact on the gas turbulence from the merger process. We only include galaxies with declination (Decl.) less than 20 degrees to be able to observe them from Siding Spring Observatory (SSO). These cuts yield a sample of 14 galaxies for which we conduct integrate field observations, of which 7 reach a signal-to-noise ratio sufficient for our further analysis (see See Section \ref{sec:2-2} and Section \ref{sec:2-3} for more details)

\begin{figure}
\includegraphics[width=\columnwidth]{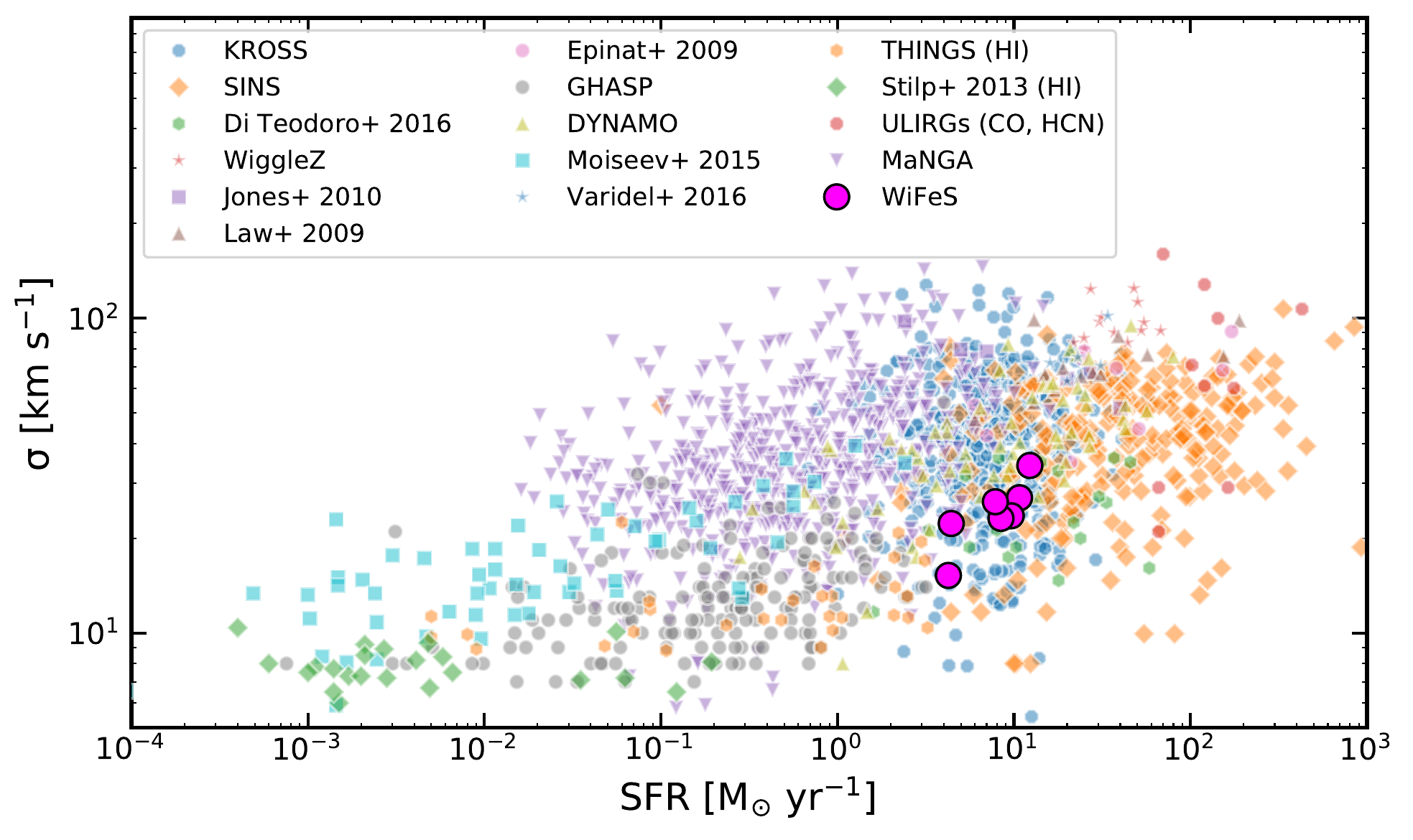}
\caption{Relationship between velocity dispersion $\sigma$ and star formation rate (SFR) in star-forming galaxies at different redshifts. The data shown include both low- and high-redshift galaxies. Except for the WiFeS galaxies (this work), other local galaxies are from \citet{Green14} (DYNAMO), \citet{Moiseev15}, \citet{Epinat08} (GHASP), \citet{Varidel16}, \citet{Yu19} (MaNGA),and  The HI of nearby galaxies are from THINGS \citep{Walter08,Leroy08,Ianjamasimanana12} and \citet{Stilp13}. The data on nearby ULIRGs are from \citet{Downes98,Sanders03,Veilleux09,Scoville15,Scoville17}. The data on high-redshift galaxies are from \citet{Law09, Epinat09,Jones10, Di16}. \citet{Wisnioski11} provided the WiggleZ sample, and \citet{Wisnioski15} and \citet{Wuyts16} provided the SINS sample. The KROSS sample is from \citet{Johnson18}.}
\label{fig:1}
\end{figure}

In Figure~\ref{fig:1}, we show the relationship between gas velocity dispersion ($\sigma$) and SFR, with galaxies drawn from the literature (both local and high redshift) shown in the background. The fuchsia foreground data points show the galaxies used in this work. Figure~\ref{fig:1} shows that there is a positive trend between $\sigma$ and SFR although with a large scatter. A potential interpretation of the large scatter in the $\sigma$ versus SFR plane, which we intend to test here, is that it is partly due to the diversity of gas fraction and rotation curve speed among the galaxies plotted.

Our sample galaxies have well measured SFR, molecular gas mass ($M_{\rm H2}$), atomic gas mass ($M_{\rm HI}$), and stellar mass ($M_*$), all of which we take from the xCOLD GASS catalog (\citealt{Saintonge17}; see details in Table~\ref{tab:1}). The SFRs are measured from WISE + GALEX for galaxies detected in both data sets, and are based on SED modeling for galaxies that lack GALEX detections \citep{Janowiecki17}. The total molecular gas mass ($M_{\rm H2}$) is derived from the total CO (1-0) line luminosity measured from the IRAM 30 meter telescope \citep{Saintonge12,Saintonge17}. The stellar mass is from the SDSS DR7 MPA/JHU catalog\footnote{\url{https://home.strw.leidenuniv.nl/~jarle/SDSS/}}, and is measured through photometry \citep{Kauffmann03,Salim07}. Finally, \citet{Catinella12} provides measurements of atomic gas mass ($M_{\rm HI}$) from observations of the H~\textsc{i} 21 cm line.

\subsection{Observations and Data Reduction}
\label{sec:2-2}
From October of 2016 to March of 2017, we observed the 14 galaxies over 6 nights using the WiFeS instrument. WiFeS is an integral field, double-beam, concentric, image-slicing spectrograph mounted on the 2.3-m telescope at SSO. It provides 25 slitlets, each 38 arcsec long and 1 arcsec wide, yielding a 25 $\times$ 38 arcsec Field of View (FoV) \citep{Dopita07,Dopita10}. The galaxies were observed using the B3000 grating on the blue arm and the R7000 grating on the red arm. The typical seeing conditions during the observations were $1.5''$ to $2''$. 

The data are reduced by the PyWiFeS data reduction pipeline \citep{Childress14}. The reduced data are reconstructed to three dimensional data cubes. In this study, we only focus our analysis on the red arm, which covers the H$\alpha$ emission line. We subtract the sky using the sky spectra extracted from the spaxels without galaxy emission. We also use the sky spectra to characterise the instrument resolution, which is crucial for the velocity dispersion estimation. We measure the velocity dispersion of four skylines that are close to H$\alpha$ emission line for each galaxy. The mean value of the measured velocity dispersion is adopted as the instrument resolution for each galaxy, which is about $\sigma_{\rm instr} \sim 17 \ \rm km \ s^{-1}$, consistent with spectral resolution $R=\lambda/\Delta\lambda$ in the WiFeS manual. We summarise the instrument resolution ($\sigma_{\rm instr}$) measurements for each galaxy in Table~\ref{tab:2}. 


\subsection{Spectral Analysis}
\label{sec:2-3}

Among the 14 galaxies selected in Section \ref{sec:2-1}, we discard one that is classified as an AGN in the xCOLD GASS catalogue \citep{Saintonge17}, and analyse the WiFeS data cubes for the remaining 13 galaxies.

\begin{figure*}
\centering
\subfigure{
\begin{minipage}[t]{0.95\linewidth}
\centering
\includegraphics[width=6.5in]{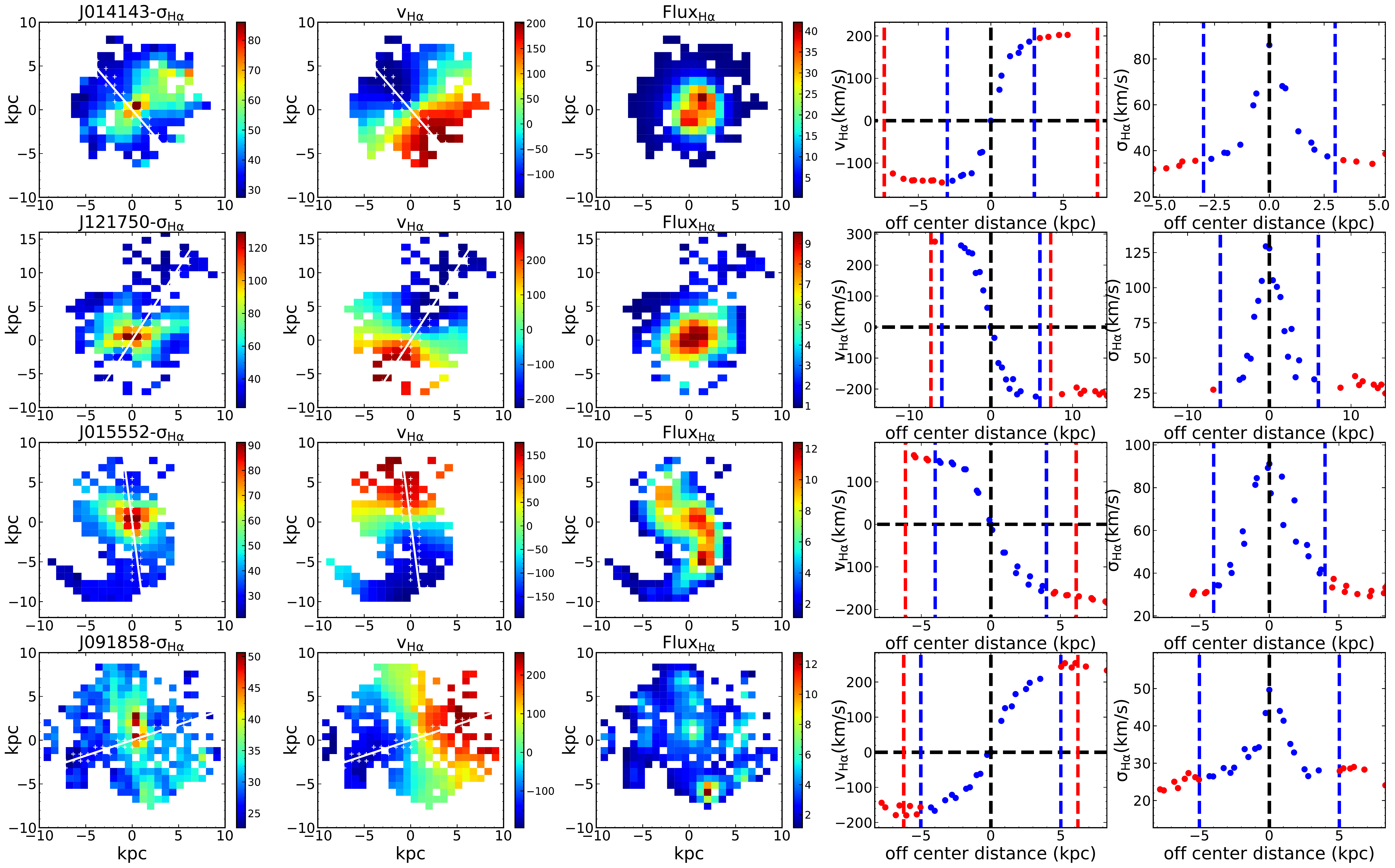}
\end{minipage}%
}%

\subfigure{
\begin{minipage}[t]{0.95\linewidth}
\centering
\includegraphics[width=6.5in]{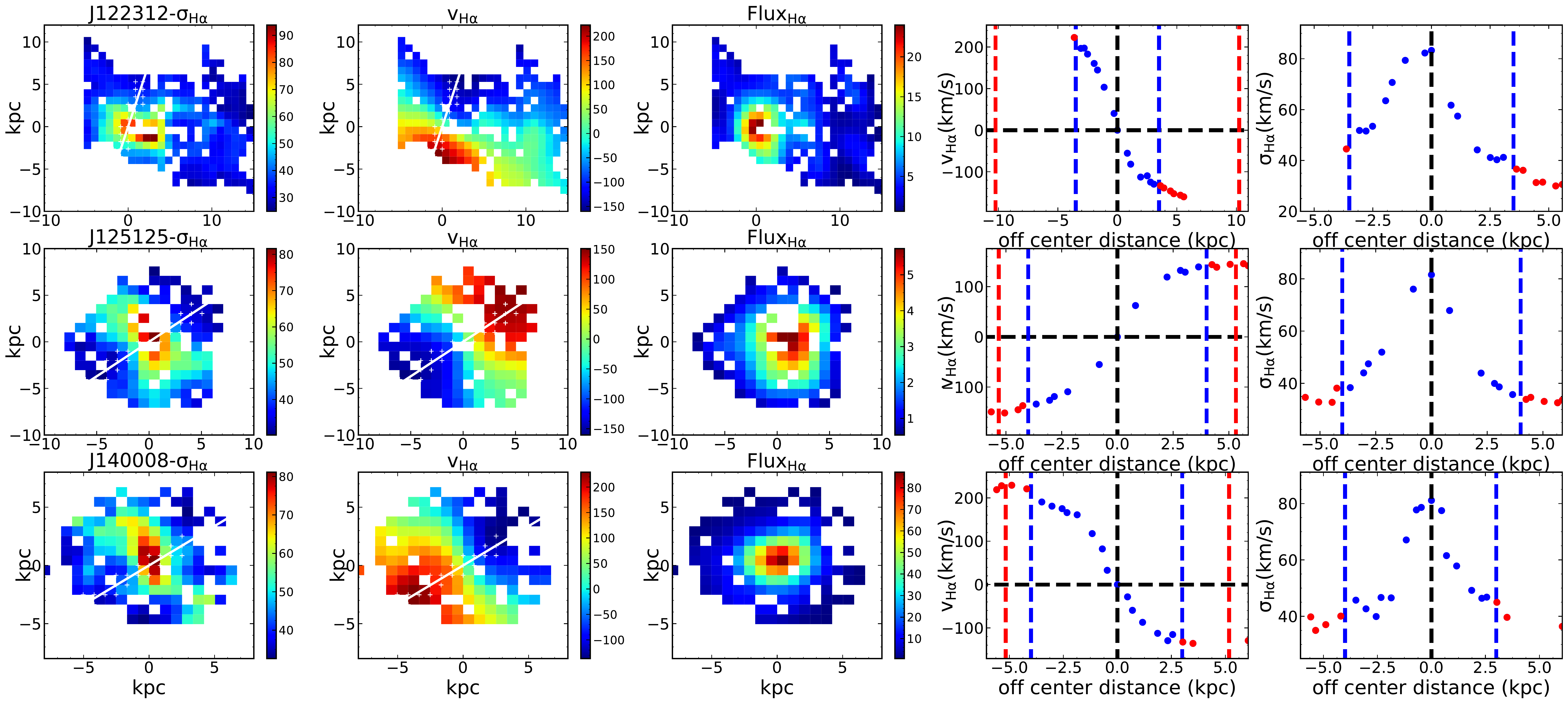}
\end{minipage}%
}%
\centering
\caption{Kinematic maps and axis profiles for the seven disc galaxies with S/N $>$ 20 that constitute our final sample. From left to right we show for each galaxy: observed H$\alpha$ velocity dispersion map, H$\alpha$ velocity field (corrected the inclination angle), H$\alpha$ emission map, velocity and velocity dispersion profile along the major kinematic axis, which is shown by the white solid lines over plotted on the velocity dispersion and velocity maps. The white cross over-plotted on the velocity dispersion and velocity maps are the data points from which we extract the velocity and velocity dispersion profiles. The blue dashed lines in the right two columns indicate the radii beyond which we deem the rotation curve to be flat; points outside the radius are shown in red, and we use these points to measure the circular velocity ($v_c$) and the observed velocity dispersion ($\sigma_{obs}$). The red dashed lines indicate the half-light radius($\rm{r_{Re}}$) we measured in section \ref{sec:2-3-2}.}
\label{fig:2}
\end{figure*}

We study galaxy kinematics using the H$\alpha$ emission line. First, the spectral continuum is subtracted using spectra at the rest-frame wavelength range between 6500 \AA \ and 6530 \AA. Then we fit the H$\alpha$, [N~\textsc{ii}]$\lambda$6549 and [N~\textsc{ii}]$\lambda$6583 emission lines simultaneously using three Gaussian profiles in each spaxel. In this process, we obtain the H$\alpha$ flux, the line of sight velocity and observed velocity dispersion ($\sigma_{\rm obs}$) for each spaxel (Figure~\ref{fig:2}).

We compute the intrinsic gas velocity dispersion for each spaxel ($\sigma_{\rm pix}$) from the observed ($\sigma_{\rm obs}$) velocity dispersion by correcting the instrument broadening as follows: $\sigma_{\rm pix} = (\sigma_{\rm obs}^2 - \sigma_{\rm instr}^2)^{1/2}$. To have a reliable ($\rm S/N > 5$) intrinsic gas velocity dispersion measurement for each spaxel, we require that the S/N of the H$\alpha$ be greater than 20 -- see the detailed discussion in \citet{Zhou17}. Seven galaxies in our sample have S/N ratios in the H$\alpha$ line that are high enough for further analysis.

\begin{table*}
\centering
\caption{Basic parameters of the final selected targets}
\label{tab:1}
\begin{tabular}{cccccccccc}
    \hline
    SDSS Name & R.A.  & Decl. & $z$ & $\log M_{*}^{\rm a}$ & $\log\mbox{SFR}^{\rm a}$ & $\log M_{\rm H2}^{\rm a}$ &  $\log M_{\rm HI}^{\rm b}$  & $f_{\rm sf}$ & $f_{g,Q}$\\
     &  &  &  & $\rm M_{\odot}$ & $\rm M_{\odot}\ yr^{-1}$ & $\rm M_{\odot}$  &$\rm M_{\odot}$ & &\\
    \hline
SDSS J014143.18+134032.8	&	01:41:43.18 	&	+13:40:32.8 	&	0.045 	&	10.67 	&	$	1.03 	\pm	0.01 	$	&	$	9.870 	\pm	0.170 	$	&	9.94 	&	$	0.46 	\pm	0.20 	$	&	$	0.52 	\pm	0.05 	$	 \\
SDSS J121750.81+082549.0	&	12:17:50.81 	&	+08:25:49.0 	&	0.049 	&	10.91 	&	$	0.64 	\pm	0.04 	$	&	$	9.457 	\pm	0.174 	$	&	10.11 	&	$	0.18 	\pm	0.07 	$	&	$	0.38 	\pm	0.01 	$ \\
SDSS J015551.98+145624.9	&	01:55:51.98 	&	+14:56:24.9 	&	0.044 	&	10.74 	&	$	0.98 	\pm	0.02 	$	&	$	9.977 	\pm	0.170 	$	&	10.22 	&	$	0.36 	\pm	0.15 	$	&	$	0.60 	\pm	0.05 	$ \\
SDSS J091858.06+055318.2	&	09:18:58.06 	&	+05:53:18.2 	&	0.038 	&	10.55 	&	$	0.63 	\pm	0.03 	$	&	$	9.729 	\pm	0.170 	$	&	9.95 	&	$	0.38 	\pm	0.16 	$	&	$	0.56 	\pm	0.04 	$ \\
SDSS J122312.26+142320.2	&	12:23:12.26 	&	+14:23:20.2 	&	0.042 	&	10.49 	&	$	0.93 	\pm	0.02 	$	&	$	9.824 	\pm	0.170 	$	&	10.33 	&	$	0.24 	\pm	0.10 	$	&	$	0.74 	\pm	0.05 	$ \\
SDSS J125125.64+035159.5	&	12:51:25.64 	&	+03:51:59.5 	&	0.049 	&	10.28 	&	$	0.89 	\pm	0.02 	$	&	$	9.731 	\pm	0.171 	$	&	9.98 	&	$	0.36 	\pm	0.15 	$	&	$	0.71 	\pm	0.07 	$ \\
SDSS J140008.99+040450.8	&	14:00:08.99 	&	+04:04:50.8 	&	0.040 	&	10.18 	&	$	1.09 	\pm	0.01 	$	&	$	9.695 	\pm	0.171 	$	&	9.69 	&	$	0.50 	\pm	0.22 	$	&	$	0.67 	\pm	0.08 	$ \\

    \hline
\end{tabular}
  
\begin{tablenotes}
\item (a): Data from xCOLD GASS catalog \citep{Saintonge17}.
\item (b): HI Data from \citet{Catinella12}.
\end{tablenotes}
      
 \end{table*}

\begin{table*}
\centering
\caption{Parameters measured from optical emission lines}
\label{tab:2}
\begin{tabular}{cccccccccccc}
    \hline
    SDSS Name & $\sigma_{\rm obs}$ &$v_c^{\rm a}$ & $\sigma_{\rm instr}$ & $\sigma_i$ & $\rm{r_{Re}}$ & $t_{\rm orb}$ & $\rm{r_{eff, H\alpha}}$ & Q & $\rm{\phi_Q}$ &$\rm Incl^{\rm b}$\\
     & $\rm km \ s^{-1}$ &$\rm km \ s^{-1}$ & $\rm km \ s^{-1}$ & $\rm km \ s^{-1}$ & $\rm kpc$ & $\rm Myr$ & kpc & & & deg\\
    \hline
       
SDSS J014143.18+134032.8	&	$	33.83 	\pm	0.30 	$	&	174.18 	&	$	16.62 	\pm	0.07 	$	&	$	29.48 	\pm	0.34 	$	&	7.35 	&	259.29 	& 5.70& 0.75 & 2.26 & 29.45\\
SDSS J121750.81+082549.0	&	$	30.69 	\pm	0.96 	$	&	248.74 	&	$	17.35 	\pm	2.74 	$	&	$	25.32 	\pm	2.21 	$	&	7.33 	&	180.95 	& 4.95& 1.04 & 3.27 & 56.36 \\
SDSS J015551.98+145624.9	&		$	31.25 	\pm	0.37 	$	&	173.37 	&	$	16.62 	\pm	0.07 	$	&	$	26.46 	\pm	0.44 	$	&	6.14 	&	217.75 &4.16 & 0.43 & 1.91 & 46.73\\
SDSS J091858.06+055318.2	&	$	26.03 	\pm	0.74 	$	&	216.87 	&	$	17.35 	\pm	1.79 	$	&	$	19.40 	\pm	1.88 	$	&	6.22 	&	176.29 	&5.72 & 0.86 & 2.08 & 18.94\\
SDSS J122312.26+142320.2	&	$	31.47 	\pm	0.29 	$	&	191.52 	&	$	17.63 	\pm	2.77 	$	&	$	26.07 	\pm	1.90 	$	&	10.24 	&	328.56 	& 3.99& 0.44 & 1.48 & 44.05\\
SDSS J125125.64+035159.5	&	$	33.09 	\pm	0.71 	$	&	153.14 	&	$	16.36 	\pm	2.44 	$	&	$	28.76 	\pm	1.61 	$	&	5.32 	&	213.24 	&4.93
&1.24 & 1.56 & 32.52\\
SDSS J140008.99+040450.8	&	$	39.64 	\pm	0.14 	$	&	182.46 	&	$	16.31 	\pm	2.46 	$	&	$	36.13 	\pm	1.12 	$	&	3.17 	&	106.84 	& 2.92 & 1.35 & 1.66 & 40.50\\

    \hline
\end{tabular}
  
\begin{tablenotes}
\item (a): Circular velocity ($v_c$) has been corrected the inclination of the disk.
\item (b): Disk inclination angle from xCOLD GASS catalog \citep{Saintonge17}.
\end{tablenotes}
      
 \end{table*}

\subsubsection{Gas velocity dispersions}
We measure the typical gas velocity dispersion for each galaxy by minimizing the ``beam smearing'' effect. The ``beam smearing'' effect blends the information from one spaxel with its neighbouring spaxels in the IFU observations due to the limited spatial resolution. Following the method proposed by \citet{Wisnioski15}, we measure velocity dispersion along the major kinematic axis from the outer regions, where the ``beam smearing'' effect is negligible. We first extract the observed velocity dispersion ($\sigma_{\rm obs}$) for each galaxy along the major kinematic axis from the observed velocity dispersion map (see Section~\ref{sec:2-3-2} for details on how we derive the major kinematic axis). We then estimate the typical $\sigma_{\rm obs}$ of our galaxies by averaging the velocity dispersion in the region of the flat rotation curve (red dots in the fifth column of Figure~\ref{fig:2}). Last, we remove the instrument resolution ($\sigma_{\rm instr}$) from the observed H$\rm{\alpha}$ gas velocity dispersion ($\sigma_{\rm obs}$) and obtain the intrinsic ionised gas velocity dispersion ($\sigma_i$) through $\sigma_i = (\sigma_{\rm obs}^2 - \sigma_{\rm instr}^2)^{1/2}$ (Table \ref{tab:2}).

Before proceeding, we must make one final correction. The measured velocity dispersions in this work are from H$\alpha$ emission lines, which are dominated by the gas from H~\textsc{ii} regions. This gas has a thermal velocity dispersion $\sigma_{\rm th} = \sqrt{k_B T/\mu m_{\rm H}}$, where $T$ is the gas temperature and $\mu$ is the mean mass of free particles, normalised to the hydrogen mass $m_{\rm H}$. This thermal velocity will be added in quadrature to the bulk velocity dispersion of the neutral ISM ($\sigma_0$) that is of interest for testing theoretical models \citep{Krumholz16,Krumholz18}. The characteristic temperature of H~\textsc{ii} regions is $T\approx 10^4$ K \citep{Andrews13}, and for fully ionised gas that is 73\% H and 25\% He by mass, $\mu=0.61$, so this corresponds to a typical thermal broadening $\sigma_{\rm th} \approx 12$ km s$^{-1}$ \citep{Zhou17}\footnote{If we were instead to assume that He is singly ionised, we would have $\mu=0.63$ and the thermal velocity dispersion would change only very slightly}. We therefore estimate the bulk ISM gas velocity dispersion $\sigma_0$ as $\sigma_0^2 = \sigma_i^2 - \sigma_{\rm th}^2$. We use $\sigma_0$ for all the analysis we present below; however, we note that, given the relatively large values of $\sigma_i$ reported in Table \ref{tab:2}, the difference between $\sigma_i$ and $\sigma_0$ is in all cases $\lesssim 20\%$.

\subsubsection{Circular Velocity and Orbital Period}
\label{sec:2-3-2}
Besides the SFR, the velocity dispersion is also predicted to depend on the circular velocity and orbital period \citep{Krumholz18}. Therefore, we also calculate these two quantities from the WiFeS IFU observations. To do so, we measure the rotation curve along the major axis using the H$\alpha$ velocity map. First, we estimate the kinematic position angle (PA) of gas and determine the major kinematic axis of each galaxy using the `FIT KINEMATIC PA' routine \citep{Krajnovic06}. The kinematic PA of gas is defined as the counter-clockwise angle between north and a line that bisects the velocity field of the gas, measured on the receding side. The first and second columns of Figure~\ref{fig:2} show the resulting fits for direction of the major kinematic axis. Then we measure the line of sight velocity along the major kinematic axis and correct the velocity for the inclination of each galaxy, using the inclination angle given in the xCOLD GASS catalog \citep{Saintonge17}. The fourth column of Figure \ref{fig:2} shows the resulting rotation curves along the major kinematic axis. We estimate the circular velocity ($v_{\rm c}$) from the flat part of the rotation curves, which we indicate by the red points in the fourth column of Figure~\ref{fig:2}. We adopt the largest value from the red dots as the circular velocity for each galaxy. We summarise the resulting values for $v_{\rm c}$ in Table \ref{tab:2}. We find that the mean circular velocity of our sample of galaxies is 191 $\rm km \ s^{-1}$.

We also measure the orbital period using the rotation curves shown in Figure~\ref{fig:2}. We first define a typical radius, half-light radius ($\rm{r_{Re}}$), so that to estimate the orbital period ($t_{\rm orb}$) for each galaxy. We used SDSS r-band image \footnote{\url{https://data.sdss.org/sas/sdsswork/atlas/}} to run GALFIT \citep{Peng02,Peng10} and obtain the half-light radius for each galaxy. One Sérsic model is enough to fit the surface brightness profile for each galaxy. The $\rm{r_{Re}}$ are listed in Table \ref{tab:2}. From the forth column of Figure \ref{fig:2}, we find that the circular velocity corresponding to half-light radius larger than the rotation curve turn-flatten radius. So when we estimate the orbital period ($t_{\rm orb}$), we used $\rm{r_{Re}}$ and $v_{\rm c}$ to estimate the $t_{\rm orb}$. We show our estimated values of $t_{\rm orb}$ in Table \ref{tab:2}. The mean value of $t_{\rm orb}$ for our galaxies is about 211 Myr.

\subsection{Theoretical models}
\label{sec:model}

\citet{Krumholz18} propose a galactic disc model based on the requirement that the disc be in both vertical hydrostatic and energy equilibrium. In this model, the sources of energy input include both star formation feedback and gravitational potential energy released by inward flow of gas through the disc. They show that these requirements lead to a relationship between star formation rate and gas velocity dispersion
\begin{eqnarray}
\rm SFR & = & 
\sqrt{\frac{2}{1+\beta}} \frac{\phi_a f_{\rm sf}}{\pi G Q} f_{g,Q} v_c^2 \sigma_0
\nonumber \\
& &
\quad {} \cdot
\max\left[
\sqrt{\frac{2(1+\beta)}{3 f_{g,P} \phi_{\rm mp}}} \frac{8 \epsilon_{\rm ff} f_{g,Q}}{Q},
\frac{t_{\rm orb}}{t_{\rm sf,max}}
\right],
\label{eq:3}
\end{eqnarray}
where $\beta$ is rotation curve index, $\phi_{a}$ is the offset between resolved and unresolved star formation law, $Q$ is the stability parameter\citep{Romeo11,Romeo13}, $\phi_{\rm mp}$ is the ratio of total pressure to turbulent pressure at mid-plane, $f_{g,P}$ and $f_{g,Q}$ are the fractional contributions of gas self-gravity to mid-plane pressure and gravitational stability, respectively, $\epsilon_{\rm ff}$ is the star formation efficiency per free-fall time, $f_{\rm sf}$ is the fraction of the ISM in the star-forming (molecular) phase, $v_c$ and $t_{\rm orb}$ are the circular velocity and orbital period at the outer edge of the star-forming disc, and $t_{\rm sf,max} = 2$ Gyr is the maximum star formation timescale. Eq. \ref{eq:3} is the prediction for a model in which both mass transport and star formation feedback contribute to energy equilibrium, and it applies for galaxies where the gas velocity dispersion is larger than $\sigma_{\rm sf}$, the maximum velocity dispersion that can be sustained by star formation feedback alone.

On the other hand, one can also use the same framework for a model in which one assumes that star formation feedback alone contributes to turbulence, and thus $\sigma_{\rm sf} = \sigma_0$ by assumption; in order for this to occur, one must allow $\epsilon_{\rm ff}$ to vary freely. Doing so yields the pure feedback model (in the terminology of \citealt{Krumholz18}, this is the feedback-only, fixed $Q$ model), for which the predicted relationship between SFR and velocity dispersion becomes
\begin{eqnarray}
\rm SFR & = & \frac{4 \eta \sqrt{\phi_{\rm mp} \phi_{\rm nt}^3} \phi_Q \phi_a}{G Q^2 \langle p_*/m_*\rangle} \frac{f_{g,Q}^2}{f_{g,P}} v_{\phi,\rm out}^2 \sigma_{0}^2.
\label{eq:4}
\end{eqnarray}
Here $\eta$ is a factor of order unity that measures the fraction of turbulent energy dissipated per crossing time, $\phi_{\rm nt}$ is the fraction of velocity dispersion that is non-thermal (close to unity for all our galaxies), $\phi_{Q}$ is the one plus ratio of gas to stellar $Q$, and $\langle p_*/m_*\rangle$ is the momentum injected per unit mass of stars formed.

More details about these models are in the work of \citet{Krumholz18}. We emphasise that, although we use the precise formulations of these models derived by \citeauthor{Krumholz18}, the predictions for the scaling between $\sigma_0$, SFR, $v_c$, and gas fraction derived there are generic to all models in the same class; for example, the scalings in Eq.~\ref{eq:4} are identical to those obtained by, e.g., \citet{Faucher-Giguere13} or \citet{Orr20}, who also present models where feedback is the sole source of turbuelence. We can therefore use the scalings in Eq.~\ref{eq:3} and \ref{eq:4} as generic tests of all models of that class.

\begin{figure*}
\includegraphics[width=5.5in]{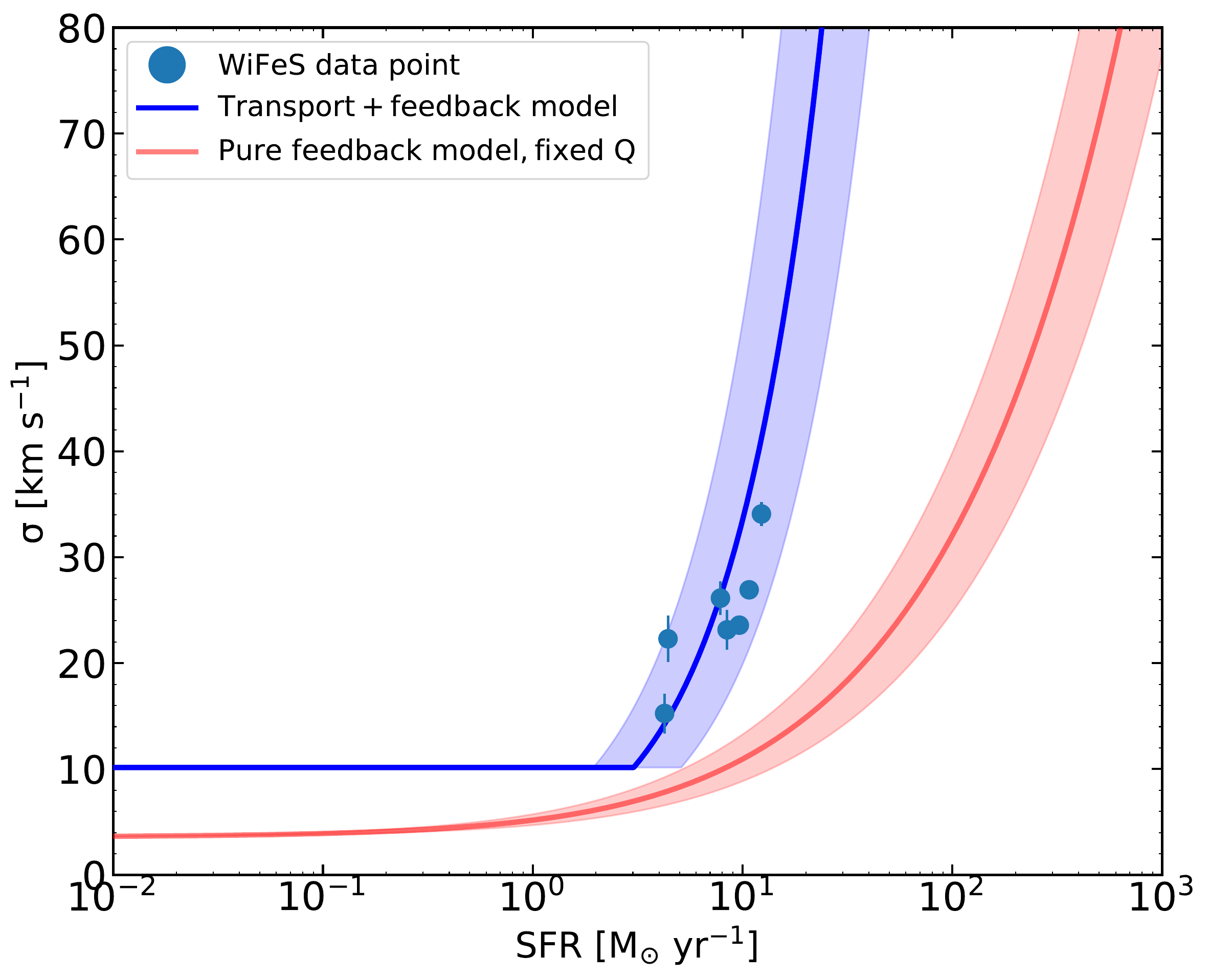}
\caption{The relationship between SFR and gas velocity dispersion. The blue and red solid lines show the transport+feedback model and pure feedback model. The shade regions represent the circular velocity range from $v_c = 153-248 \ \rm km \ s^{-1}$, which is the typical range for our galaxies. The $\sigma$ we use in this work has been subtracted the thermal broadening in quadrature of HII region, about $12 \ \rm km \ s^{-1}$. This plot shows that the transport+feedback model are agreement with the observations for our galaxies and transport dominates the high $z$ and high star-forming galaxies}
\label{fig:3}
\end{figure*}


\section{Results and Discussion}
\label{sec:5}
In this section, we compare our observational results with the theoretical models described in Section \ref{sec:model} and discuss the results.

\begin{figure*}
\includegraphics[width=7.0in]{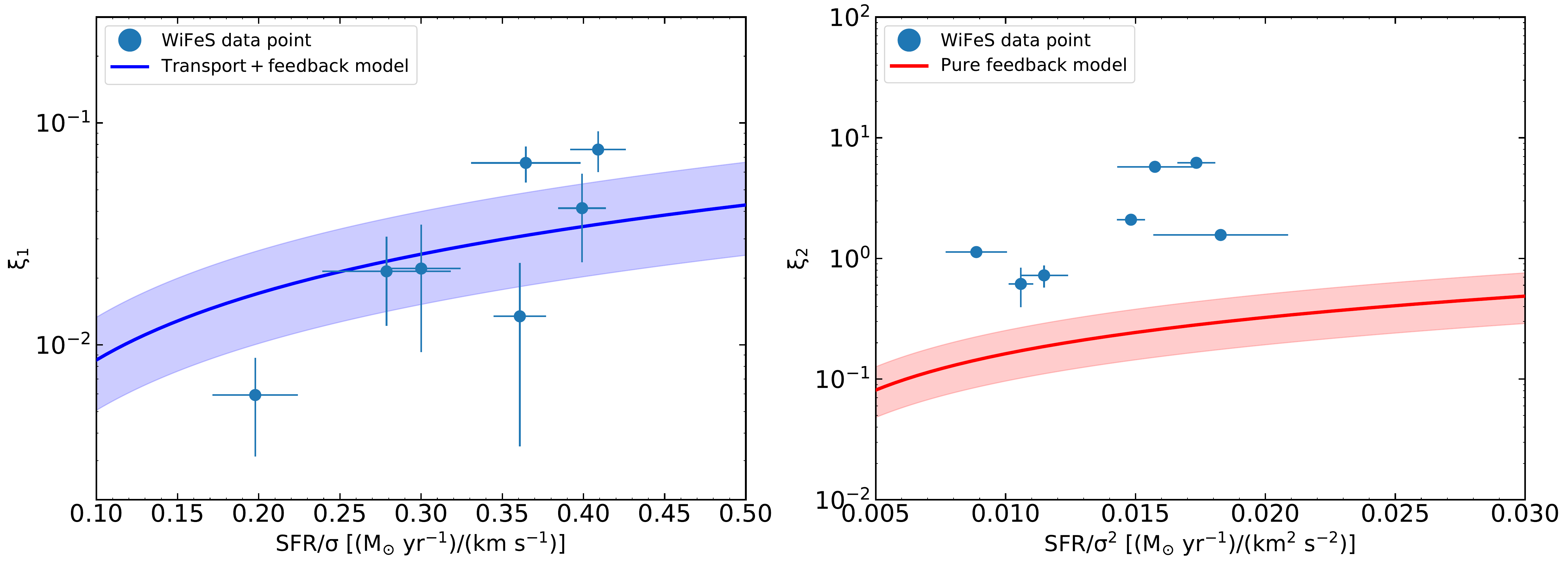}
\caption{Left: The relationship between $\rm{\xi_1}$ and the ratio of SFR to velocity dispersion ($\rm{SFR/\sigma}$). The blue solid lines shows the transport+feedback model. Right: The relationship between $\rm{\xi_2}$ and the ratio of SFR to the square of velocity dispersion ($\rm{SFR/\sigma^2}$). The red solid lines represent the pure feedback model. Both left and right panels, the shade regions represent the circular velocity range from $v_c = 153-248 \ \rm km \ s^{-1}$, which is the typical range for our galaxies. The $\sigma$ we use in this work has been subtracted the thermal broadening in quadrature of HII region, about $12 \ \rm km \ s^{-1}$.}
\label{fig:4}
\end{figure*}

\subsection{Measurement of the model input parameters}
\label{sec:4-1}

Our IFU observations provide straightforward measurements of some of the parameters that enter the two theoretical models presented in Section \ref{sec:model}: the gas velocity dispersion ($\sigma_0$), orbital period ($t_{\rm orb}$), and circular velocity ($v_c$). Similarly, the xCOLD GASS survey provides the SFR and gas content in these galaxies, which allow us to estimate the gas fraction in the star-forming molecular phase as $f_{\rm sf} = (M_{\rm H2})/(M_{\rm H2} + M_{\rm HI})$. We report our measured values of $f_{\rm sf}$ in Table \ref{tab:1}; the mean value of $f_{\rm sf}$ is about 0.35. 

Other parameters appearing in the theoretical models require somewhat more indirect estimation procedures. Following \citet{Krumholz18}, we estimate $f_{g,P}\approx f_{g,Q}$, and we compute $f_{g,Q}$, the effective gas fraction in the disc for the purpose of computing gravitational stability, from its definition
\begin{equation}
\label{eq:fgq}
f_{g,Q}\equiv \frac{\Sigma_g}{\Sigma_g + [2\sigma_0^2/(\sigma_0^2+\sigma_*^2)]\Sigma_*},
\end{equation}
where $\Sigma_g$ is gas surface density, $\Sigma_*$ is stellar surface density, $\sigma_g$ is the intrinsic gas velocity dispersion, and $\sigma_*$ is the stellar velocity dispersion. We lack direct measurements of $\sigma_*$, and we therefore adopt the ratio $\sigma_* / \sigma_0 = 2.3$ measured in the Milky Way \citep{Kalberla09,McKee15,Krumholz18}. This ratio may well be somewhat smaller in our more gas-rich galaxies, which presumably have dynamically ``hotter'' gas discs. However, we have verified that even using an extreme value $\sigma_*/\sigma_0 = 1$ (i.e., gas and stars have the same velocity dispersion) does not make a substantial difference to our results. For this reason we simply adopt the Milky Way value for the remainder of the paper. Similarly, we do not have independent measures of the stellar and gas areas, and we therefore adopt $\Sigma_{\rm g}/\Sigma_* = (M_{\rm H2} + M_{\rm HI})/M_*$. Table \ref{tab:1} shows our measured values of $f_{g,Q}$. The mean value is about 0.60, slightly higher than for the Milky Way ($f_{g,Q}\approx 0.5$ -- \citealt{Krumholz18}), which is not surprising given that we have intentionally selected more gas-rich galaxies. 

Similarly, we estimate the stability parameters for our discs from
\begin{equation}
\label{eq:Q}
Q = f_{g,Q} Q_{\rm g}, 
\end{equation}
where $Q_{g} = \kappa \sigma_0/\pi G \Sigma_{g}$. Here, $\kappa = \sqrt{2(\beta+1)}\Omega$ is the epicyclic frequency  and $\Omega = 2\pi/t_{\rm orb}$ is the galaxy angular velocity \citep{Krumholz18}. We can measure $\Omega$ directly from the orbital period, and all our galaxies have relatively flat rotation curves outside of their central regions, so we adopt $\beta = 0$, which provides an estimate of $\kappa$. The gas surface density, $\Sigma_g$, requires somewhat more care because xCOLD GASS does not resolve the target galaxies. For the H~\textsc{i} component, we rely on the well-established observational and theoretical result galactic H~\textsc{i} surface densities never exceed $\approx 10$ M$_\odot$ pc$^{-2}$, because higher gas column densities lead to the formation of self-shielded regions that convert to H$_2$ \citep{Krumholz08,Krumholz09,Kalberla09,McKee10,Wong13}. Since our galaxies are gas-rich and more rapidly star-forming than the Milky Way, we expect them to lie near this limit, and adopt $\Sigma_{\rm HI} = 10$ M$_\odot$ pc$^{-2}$ as the surface density of the atomic component. For the molecular gas, we assume that the H$\alpha$ emission traces star-forming molecular gas, consistent with the observation that H$\alpha$ and CO  extremely well-correlated when averaged on $\gtrsim$ kpc scales (e.g., \citealt{Leroy08, Leroy13a}). We therefore use the resolved H$\alpha$ emission as a proxy for the radius of the unresolved CO. We compute a third-order polynomial to fit the spatial $\rm{H\alpha}$ flux as a function of radius, and use this fit to obtain the half-light radius of ionized gas emission ($r_{\rm eff,H\alpha}$), which we list in the Table \ref{tab:2}, and derive the surface density of H$_2$ from $M_{\rm H2}/\pi r_{\rm eff,H\alpha}^2$. Inserting these values into Eq.~\ref{eq:fgq} yields estimates of $Q$ for each galaxy, which we report Table \ref{tab:2}. The mean value of Q is 0.77.

The final parameter we estimate from observations is 
\begin{equation}
\label{eq:phiQ}
\phi_Q \equiv 1 + \frac{Q_{\rm g}}{Q_*},
\end{equation}
where $Q_* = \kappa \sigma_*/\pi G \Sigma_*$. We can estimate $Q_*$, and thence $\phi_Q$, using the same assumed ratio $\sigma_*/\sigma_0$ adopted above. Re-arranging the definitions of $Q$, $Q_*$, and $f_{g,Q}$, we have
\begin{equation}
\label{eq:q}
Q \approx \left(Q_g^{-1} + \frac{2\sigma_{\rm g}\sigma_*}{\sigma_{\rm g}^2+\sigma_*^2} Q_*^{-1}\right)^{-1},
\end{equation}
and plugging our estimates of $Q$\citep{Krumholz18,Romeo11,Romeo13}, $Q_g = f_{g,q} Q$, and $\sigma_*/\sigma_0$ into this expression yields $Q_*$, and then subsituting into Eq. \ref{eq:phiQ} gives $\phi_Q$. We report our estimate for this parameter in Table \ref{tab:2}. The mean value is ${\phi_Q}$ is 2.03.

For all the remaining parameters that enter Eq.~\ref{eq:3} and Eq.~\ref{eq:4}, we adopt the fiducial values of \citet{Krumholz18}. We list these for convenience in Table \ref{tab:par}.

\begin{table}
\centering
\caption{Fiducial parameter values from \citet{Krumholz18}}
\begin{tabular}{ccccccc}
\hline
Parameter & Value \\
\hline
$\rm{t_{sf, max}}$ \ [Gyr] & 2.0 \\
$\rm{\beta}$  & 0 \\
$\rm{\phi_{mp}}$ & 1.4 \\
$\rm{\epsilon_{ff}}$ & 0.015\\
$\rm{\eta}$ & 1.5\\
$\rm{\phi_{nt}}$ & 1.0\\
$\rm{\phi_{a}}$ & 3.0\\
$\rm{p_{*}/m_{*} \ [km \ s ^{-1}]}$ & 3000\\
\hline
\end{tabular}
\label{tab:par}
\end{table}

\subsection{Comparison of theoretical models to observations}

\label{sec:5-2}
We are now in a position to compare our observational results with the theoretical models discussed above: the transport+feedback model (Eq.~\ref{eq:3}) that includes both gravitationally-driven and feedback-driven contributions, and the feedback-only model (Eq.~\ref{eq:4}). 

Figure \ref{fig:3} shows the relationship between SFR and the intrinsic gas velocity dispersion. The solid blue line and solid red lines show the transport+feedback model and pure feedback model, respectively. For the purposes of generating the theoretical lines, we adopt the mean values of measured parameters, which are $f_{\rm sf} = 0.35$, $v_c = 191 \ \rm km \ s^{-1}$, $t_{\rm orb} = 211 \ \rm Myr$, $Q=0.77$, and $f_{g,Q}=f_{g,P}=0.60$ (see Table \ref{tab:1}, Table \ref{tab:2}, and Section \ref{sec:2-3-2}). The blue and red shaded regions represent the range from $v_c = 153-248 \ \rm km \ s^{-1}$, which is the range of measured circular velocity in our galaxies. We find that the transport+feedback model prediction is in better agreement with the observations for our galaxies. The stellar feedback only model significantly underestimates the velocity dispersion for a given SFR.  

Because we have individual measurements for the gas fractions, we can use also these values instead of the means, thereby performing a test of the ``secondary'' dependence of the SFR on gas fraction (as opposed to the ``primary'' dependence on velocity dispersion). To this end, we re-arrange Eq.~\ref{eq:3} so as to isolate the dependence on gas fraction. We therefore define the quantity
\begin{equation}
    \xi_1 \equiv \frac{f_{\rm sf} f_{g,Q}}{Q} \max\left[
\sqrt{\frac{2(1+\beta)}{3 f_{g,P} \phi_{\rm mp}}} \frac{8 \epsilon_{\rm ff} f_{g,Q}}{Q},
\frac{t_{\rm orb,out}}{t_{\rm sf,max}},
\right].
\end{equation}
which captures the dependence of the model prediction on gas fraction (and Toomre $Q$). Examining Eq.~\ref{eq:3}, we can see that, in the transport+feedback model, we expect $\mbox{SFR}/\sigma \propto \xi_1$. Performing an analogous operation for the pure feedback model (Eq.~\ref{eq:4}), we find that the gas fraction dependence is captured by the quantity
\begin{equation}
    \xi_2 = \frac{f_{g,Q}\phi_{Q}}{Q^2}.
\end{equation}
Again, comparison to Eq.~\ref{eq:4} shows that, in the pure feedback model, we expect $\mbox{SFR}/\sigma^2\propto \xi_2$.  We plot these two predictions against the data in the left and right panels of Figure \ref{fig:4}, respectively. First examining the transport+feedback model, we find that, given the observational uncertainties, the model prediction agrees well with the data, both in normalisation and in trend. We emphasise that this is a non-trivial result: the model predicts that, at fixed star formation rate, more gas-rich galaxies (higher $\xi_1$) tend to have smaller velocity dispersion, and the data bear out this prediction. By contrast the feedback-only model in the right panel does not match the data in normalisation, which is not surprising, since we have already seen this effect in Figure \ref{fig:3}. While the trend in $\xi_2$ with $\mbox{SFR}/\sigma^2$ is qualitatively consistent with the prediction within the errors, the observed galaxies are much more gas-rich (higher $\xi_2$) at a given value of $\mbox{SFR}/\sigma^2$ than would be expected for the feedback-only model.

\subsection{Caveats}
\subsubsection{Sample selection}
As discussed in Section \ref{sec:2-1}, our sample selection is largely driven by the need to explore parts of parameter space where feedback-only and feedback-plus-gravity models make measurably different predictions, as well as where we also have access to data on the gas content of galaxies. These two goals are in tension, because the predictions of the two models differ most strongly for high-SFR, highly-turbulent galaxies of the type found primarily at high-redshift, but for such galaxies we have little information on their molecular content and essentially none on their H~\textsc{i} content. Conversely, our knowledge of gas properties is the richest for nearby galaxies, but the vast majority of these fall into the low-SFR, low-turbulence regime where the feedback-only and feedback-plus-gravity models make nearly identical predictions. As a result of these competing imperatives, our sample winds up covering a fairly narrow range of SFR and velocity dispersion. While our exploration of this regime favours the feedback-plus-gravity model, we cannot from this sample alone rule out the possibility that a feedback-only model might provide a better fit to data in other parts of parameter space, for example at SFRs $\gtrsim 30$ M$_\odot$ yr$^{-1}$ or velocity dispersions $\gtrsim 40$ km s$^{-1}$, larger than those we have been able to explore.

\subsubsection{Bias in H$\alpha$ velocity dispersions}

A second caveat to our work concerns our use of H$\alpha$, which traces ionised gas, as our tool for measuring the velocity dispersion. In deriving the velocity dispersion of the dominant neutral ISM component, we have corrected for thermal broadening, but this correction is relatively small ($\approx 20\%$) for the range of velocity dispersions found in our sample galaxies. Our approach is consistent with the observational results of \citet{Ubler18} and \citet{Girard19}, who find that the ionised gas velocity dispersion differs little from the molecular gas velocity dispersion in galaxies where the total velocity dispersion $\sigma \gtrsim 20$ km s$^{-1}$. More recently, however, \citet{Girard21} obtained much larger differences between molecular and ionised gas velocity dispersions, a factor of $\approx 2.5$, in a sample of 9 local galaxies, several with velocity dispersions similar to the range we are exploring, $\gtrsim 20$ km s$^{-1}$. These results are inconsistent with the previous findings in the literature, and the origin of the discrepancy is unknown. Further complicating this story, recent simulations by \citet{Kretschmer21} suggest that galaxies with large velocity dispersions can show a significant systematic offset in velocity dispersion between molecular and \textit{atomic} phases of the ISM, such that the ionised gas velocity disperison is actually closer than the molecular gas value to the velocity dispersion of the neutral ISM as a whole.

Given the divergent and contradictory results in the literature, we have elected to adopt a minimal approach of not making a larger correction to the H$\alpha$ velocity dispersion. However, it is important to consider how such a larger correction would affect our results. As an extreme, adopting a simple factor of $2.5$ reduction to all our velocity dispersions, consistent with \citet{Girard21}, would have the effect of bringing our measurements much more closely into alignment with the feedback-only model, and pushing them further from the feedback-plus-gravity one. However, such a change would also have other effects; for example, such a correction would reduce our estimated values of Toomre $Q$ to $\approx 0.3$, which would likely invalidate the feedback-only model as well, since even in this model it is assumed that galaxies self-regulate to $Q\approx 1$.

\section{Conclusions}
\label{sec:6}
We summarise our main results are as follows:
\begin{enumerate}
\item We select a sample of star-forming galaxies in the xCOLD GASS survey, chosen to lie in a range of star formation rate and the gas fraction that offers maximum power to distinguish between alternative models for the origin of turbulence in the ISM: on where stellar feedback alone provides the required energy, and where both stellar and gravitational instability-driven transport of mass down the potential well supply energy to the turbulence.
\item We carry out IFU observations of these galaxies with the WiFeS instrument, obtaining robust measurements of the gas velocity dispersion, circular velocity, and orbital period in our galaxies.  These parameters, together the gas fractions taken from the xCOLD GASS survey, provide the inputs required by the competing theoretical models.
\item We compare the relation between SFR, velocity dispersion, and gas fraction with the two models. We find that our results are consistent with the transport+feedback model. In particular, not only does this model match the mean relationship between velocity dispersion and star formation rate seen in the data, it reproduces the secondary trend with gas fraction. By contrast, the feedback-only model both underpredicts the velocity dispersion and fails to match the dependence on gas fraction. 
\end{enumerate}

Overall, we find that the transport+feedback model is a better match to the observations. This suggests that stellar feedback only cannot provide the energy to support the high velocity dispersion found in our galaxies. Our results suggest that the gravitational instability dominates the high turbulence of high $z$ and high SFR star-forming galaxies.

\section*{Acknowledgements}
We thank the referee for a detailed report that helped significantly in improving the presentation of our work. X.Y. and Y.S. acknowledge the support from the National Key R\&D Program of China (No. 2017YFA0402704, No. 2018YFA0404502), the National Natural Science Foundation of China (NSFC grants 11825302, 11733002 and 11773013). MRK acknowledges function from the Australian Research Council through its \textit{Discovery Projects} and \textit{Future Fellowship} schemes (awards DP190101258 and FT180100375). FB acknowledges support from the Australian Research Council through Discovery Projects (award DP190100252) and Chinese Academy of Sciences (CAS)
through a China-Chile Joint Research Fund (CCJRF1809) administered by the CAS South America Center for Astronomy (CASSACA).

\section*{DATA AVAILABILITY}
The data underlying this article will be shared on reasonable request to the corresponding author.






\bibliography{reference}



%
%


\bsp	
\label{lastpage}
\end{document}